\documentclass{article}

\usepackage{natbib}
\usepackage{graphicx}
\usepackage{psfrag}

\title{Instability of stretched and twisted soap films in a cylinder}

\author{Simon Cox and Si\^an Jones\footnote{Present address: Laboratoire des Physique des Solides, Universit\'e Paris Sud, 91405 Orsay, France.}\\
Institute of Mathematics and Physics, Aberystwyth University, SY23 3BZ, UK}

\date{March 2013}

\begin{document}

\maketitle

\begin{abstract}
A soap film, or a flexible membrane without bending and torsional stiffness, that is confined in a cylinder is shown to be susceptible to a surface-tension-driven instability when it is stretched or twisted. This leads to its breakdown and places an upper limit on the aspect ratio of such structures. A simple analysis confirms the values for the critical aspect ratio of the stretched film found in both simulations and experiments on soap films, and this threshold decreases with increasing twist of the film.
\end{abstract}


\section{Introduction}

Area-minimizing interfaces, such as soap films and elastic membranes 
without bending and torsional stiffness, are found in a wide variety of 
man-made and biological structures, and in processes such as liquid 
confinement in microgravity and enhanced oil recovery.  The ability to 
predict their shape, strength and/or integrity is therefore important.

That soap films are not indefinitely stable is clear from the well-known 
case of the catenoidal film formed between two closely-spaced parallel rings 
of radius $R$
\cite{isenberg92}. As the two rings are pulled apart, the neck 
gradually thins. This does not continue indefinitely, and at a critical 
ring separation $H \approx 1.32R$ the surface becomes unstable and jumps to two discoidal films, one in each 
ring.

A related instability is the stretching of a cylindrical bubble between 
parallel walls \cite{isenberg92}. Above a certain wall separation, $H = 2 \pi R$, the bubble 
 undergoes a ``wine-bottle'' \cite{coxwv01}, or Rayleigh-Plateau, instability, bulging at one end and 
thinning at the other. As the stretching continues, the bubble detaches from one wall, 
and takes a hemispherical shape \cite{WeaireVTF07}.



Other soap-bubble instabilities are reviewed by \citeauthor{WeaireVTF07} (\citeyear{WeaireVTF07}). 
They make the implicit assumption, as we shall do throughout the 
following, that the amount of surfactant present 
is sufficiently great that a film does not collapse because its tension 
becomes too high and that instabilities are not driven by gradients in surface tension. 
This assumption is certainly valid in our experiments, 
described below.

We first consider the stability of a single planar film. A single 
rectangular soap film confined between parallel walls, with its ends held by two 
wires perpendicular to the walls, can be stretched 
indefinitely by pulling the anchoring wires apart.

However, if the film is 
confined in a cylinder, oriented parallel to the axis and such that the 
anchoring wires are parallel diameters of the cylinder,
we find that this cylindrical constraint induces a surface-tension-driven 
instability in sufficiently long films.
Once this instability is triggered, at some critical aspect ratio, the film 
rapidly deforms and collapses. If the film is twisted, taking the shape of a helicoid, 
then the instability occurs when the aspect ratio is lower.

\begin{figure}
\centerline{
\includegraphics[width=\textwidth]{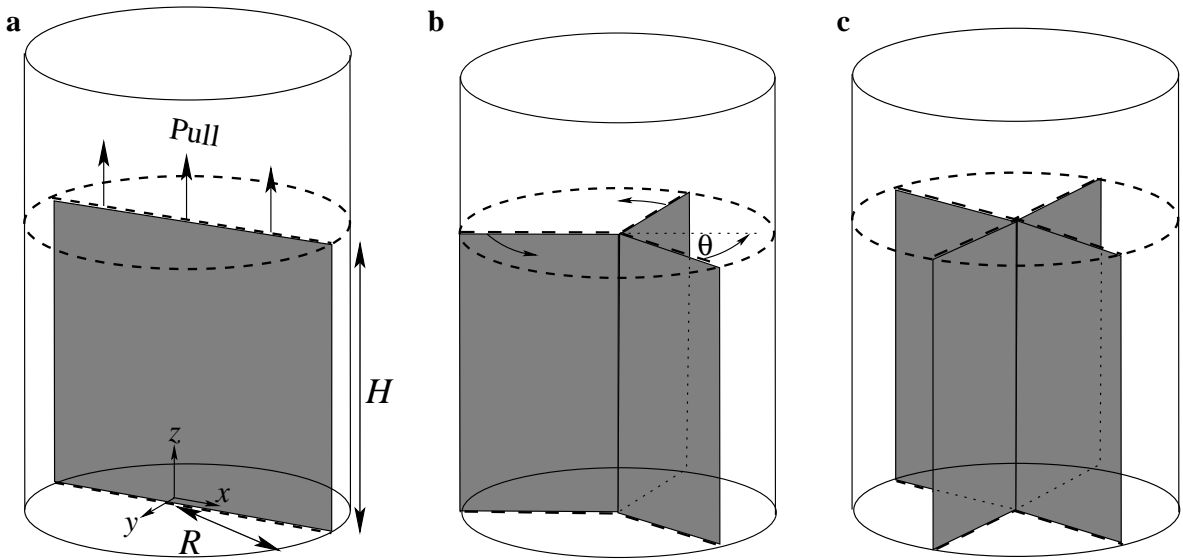}
}
\caption{Setup. (a) A rectangular film confined within a cylinder. The two wires on which the film is formed are shown as dashed lines. The film is fixed to these wires, but free to move on the cylinder wall. The two wires are pulled apart along the axis of the cylinder (as shown) or one is twisted relative to the other. 
(b) Arrangement of three films confined within a cylinder. The films are supported by $Y$-shaped wires at each end, meet each other at $120^\circ$, and are again free to move on the cylinder wall. The upper Y-shaped arrangement of anchoring wires is rotated at fixed aspect ratio, or pulled upwards at fixed $\theta$ (not shown). 
(c) Arrangement of four elastic membranes in a cylinder (simulation only).}
\label{fig:one}
\end{figure}

Most physical systems, however, consist of more than one soap film. In more complex arrangements of 
multiple soap films, a consequence of area minimization is that these 
films meet in threes at equal angles of $120^\circ$ 
\cite{plateau73,taylor76} in what are known as Plateau borders. Thus, it is important in predicting the 
stability of foams, which are used in many processes including froth 
flotation and ore separation \cite{WeaireH99,mousse10}, to determine the 
degree to which this three-fold arrangement of soap films is susceptible 
to the same surface-tension-driven instability.  We therefore add more 
soap films parallel to the axis of the cylinder, in both simulation and 
experiment, and show that these configurations confer no greater stability 
on the system. The three systems investigated here are shown in Figure \ref{fig:one}.

\section{Experiments}

\begin{figure}
\centerline{
\includegraphics[width=0.35\textwidth]{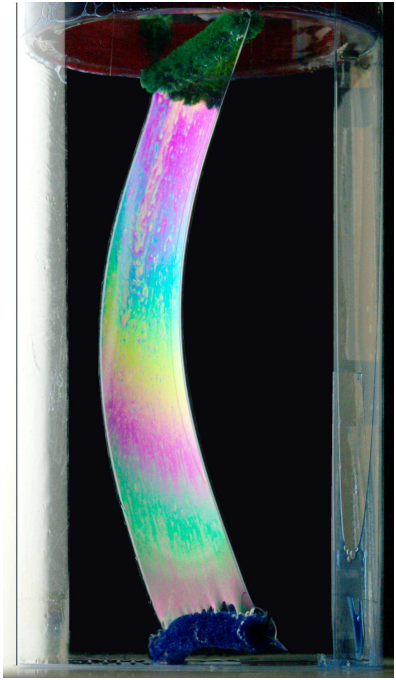}
}
\caption{An experiment on a single stretched film is shown just after the instability is triggered. The film bows outwards as it collapses.
}
\label{fig:expt}
\end{figure}

We used a 10\% solution of the commercial detergent Fairy Liquid to create 
a single soap film between two parallel pieces of pipe-cleaner (referred to here as wires) placed  
perpendicular to the axis of a perspex cylinder of fixed radius $R=31$mm and length 300 mm (as in Figure  \ref{fig:one}(a)). 
The upper wire is held in place with a thin perspex ring attached to a handle that allows its position to be controlled.
Note that there is no volume constraint on the gas neighbouring the film.

The film was then slowly stretched by increasing the distance $H$ between the 
wires in small steps from zero (the point at which the film is formed).
We define the aspect ratio of the film to be $\alpha = H/R$ (rather 
than the more conventional definition, which is half this) for consistency 
with the experiments described below with multiple films.  The critical 
aspect ratio at which the film becomes unstable was found to be $\alpha_{\rm crit}=2.76 \pm 0.15$.

The instability becomes apparent when the film starts to bow outwards while remaining attached to the wires at top and bottom, as shown in Figure \ref{fig:expt}.
The midpoints of the vertical sides of the film move around the cylinder wall until they meet. The film then pinches off into two parts, each of which retract towards the ends of the cylinder. The final state consists of two semicircular films between a wire and the cylinder wall, one at each end of 
the cylinder. 
This whole process, from initial bowing of the film to the final collapsed state, can take several tens of seconds due to friction at the wall.  

\begin{figure}
\psfrag{Twist angle [degrees]}{Twist angle [degrees]}
\psfrag{Aspect ratio [a = H/R]}{Aspect ratio [$\alpha = H/R$]}
\centerline{
\includegraphics[width=0.9\textwidth]{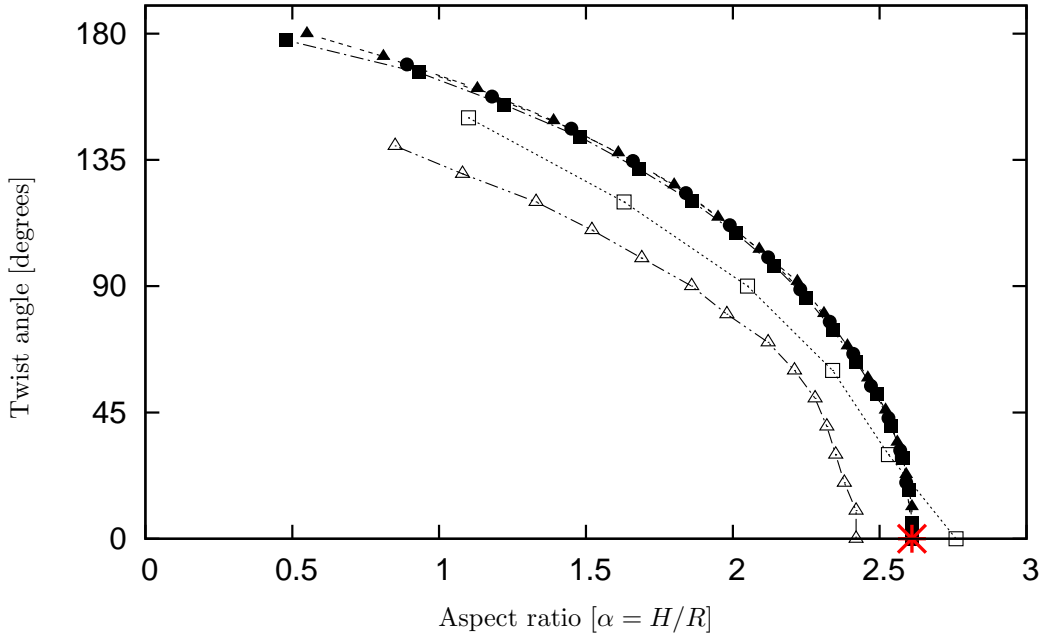}
}
\caption{The critical aspect ratio necessary to trigger the instability decreases with increasing twist angle, or, equivalently, the critical twist angle decreases with increasing aspect ratio. Key: simulations are shown with filled symbols, experiments with open symbols; single film: squares, three-fold film arrangement: triangles, and four-fold film arrangement: circles. Measurement errors are not shown, for clarity, but are estimated at: experiment $\pm 0.15$ for aspect ratio, $\pm 10^\circ$ for angle; simulation $\pm 0.01$ for aspect ratio, $\pm 0.01^\circ$ for angle. The theoretical prediction at $\alpha = 2.619$, $\theta = 0$ is shown as an asterisk on the horizontal axis.}
\label{fig:plot}
\end{figure}

We next formed a film and stretched it to a small value of $H$, then twisted it into a helicoid by slowly moving the upper constraint clockwise up to an angle $\theta$. We then again stretched the film by increasing $H$ and recorded the value of $H/R$ at which the film became unstable and collapsed. The critical aspect ratio {\em decreased} with increasing $\theta$ in the manner shown in Figure \ref{fig:plot}. Equivalently, at given aspect ratio $\alpha$, the film can only be twisted up to a certain angle until it becomes unstable and then collapses as before.

Both stretching and twisting  experiments were repeated with a Y-shaped frame, again made from pipe-cleaners (see Figure  \ref{fig:one}(b)), to generate data for the collapse of the three-fold film arrangement. Each film is a helicoid, and the instability is first apparent when the triple line along the axis of the cylinder (i.e. the Plateau border) deforms.  The critical values of the aspect ratio and twist angle are also shown in Figure  \ref{fig:plot}: they are close to those for the single film case, suggesting that the same mechanism for instability is involved, but are at slightly lower aspect ratio for given twist angle.

\section{Simulations} 

We used the Surface Evolver \cite{brakke92} to determine the shape of a single film at a minimum of surface area. We set $R=1$ and then tessellated the film with about 2000 triangles (three levels of refinement). The edges that lie along the diameters at the end of the cylinder were fixed, and the sides of the film were constrained to move on the curved walls of the cylinder. On stretching the film, in steps of $\delta \alpha = 0.01$, we found that an eigenvalue of the Hessian of energy \cite{brakke96} first become negative at $\alpha_{\rm crit} = 2.61 \pm 0.01$, close to the value found in the experiments. 

\begin{figure}
\centerline{
\hspace{0.05\textwidth}
 {\bf  \small a}
\hspace{0.35\textwidth}
 {\bf  \small b}
\hspace{0.40\textwidth}
}
\centerline{
\includegraphics[width=0.35\textwidth]{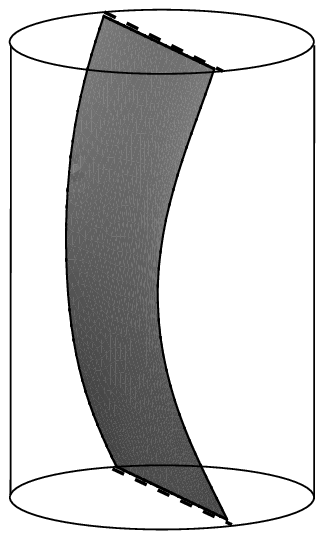}
\includegraphics[width=0.40\textwidth]{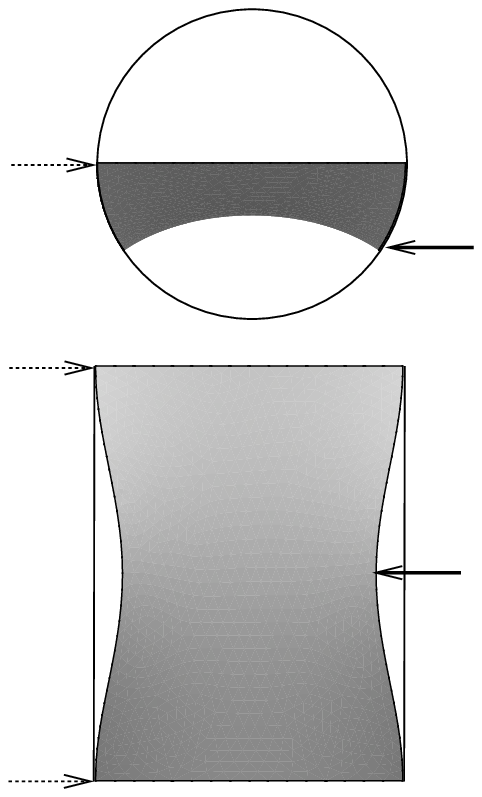}
}
\centerline{
\hspace{0.05\textwidth}
 {\bf  \small c}
\hspace{0.4\textwidth}
 {\bf  \small d}
\hspace{0.3\textwidth}
}
\centerline{
\includegraphics[width=0.3\textwidth]{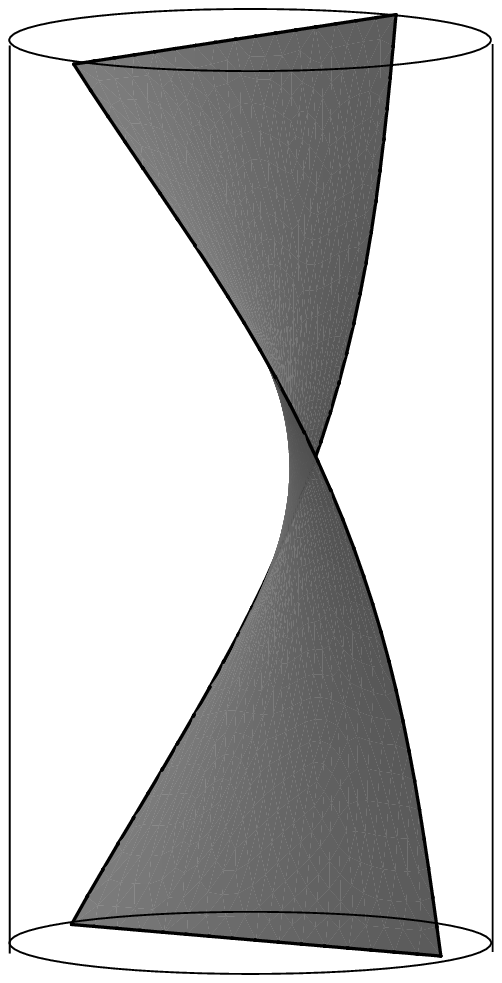}
\hspace{0.1\textwidth}
\includegraphics[width=0.3\textwidth]{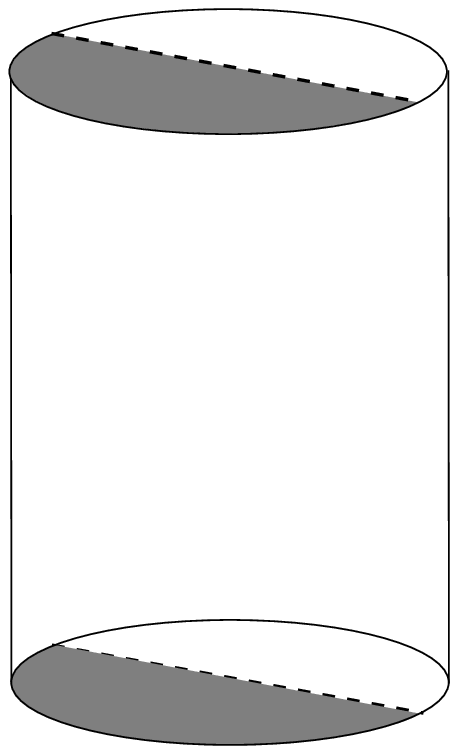}
}
\caption{Shapes of a single film after the instability is initiated. (a) After stretching: the centre of the film bulges out and appears to be pinched where it curves around the cylinder. (b) View of the film in (a) from in front and above. The arrows identify the position of the top and bottom of the film (left) and the midpoint (right) in each view.  (c) After twisting: similar to (a). (d) The final collapsed state: two semi-circular films.}
\label{fig:two}
\end{figure}

Starting from a small value of $H$, we then twisted the film in increments of one-tenth of a radian before stretching, as in the experiments. In other words, we searched for the instability by increasing $\alpha$ at fixed $\theta$; this turns out to be equivalent to fixing $\alpha$ and increasing $\theta$. We again recorded the value of $\alpha$ at which an eigenvalue of the Hessian became negative, as a measure of the triggering of the instability. The critical value of $\alpha$ decreased with increasing twist angle, as shown in Figure \ref{fig:plot} and in broad agreement with the experiments.

Continuing the simulation after the instability occurs allowed us to produce illustrative snapshots of the collapse of the film (Figure \ref{fig:two}), although these should be viewed with caution because we have not attempted to model the dynamics of the film.

\begin{figure}
\centerline{
\hspace{0.05\textwidth}
 {\bf \small a}
\hspace{0.35\textwidth}
 {\bf \small b}
\hspace{0.35\textwidth}
}
\centerline{
\includegraphics[width=0.35\textwidth]{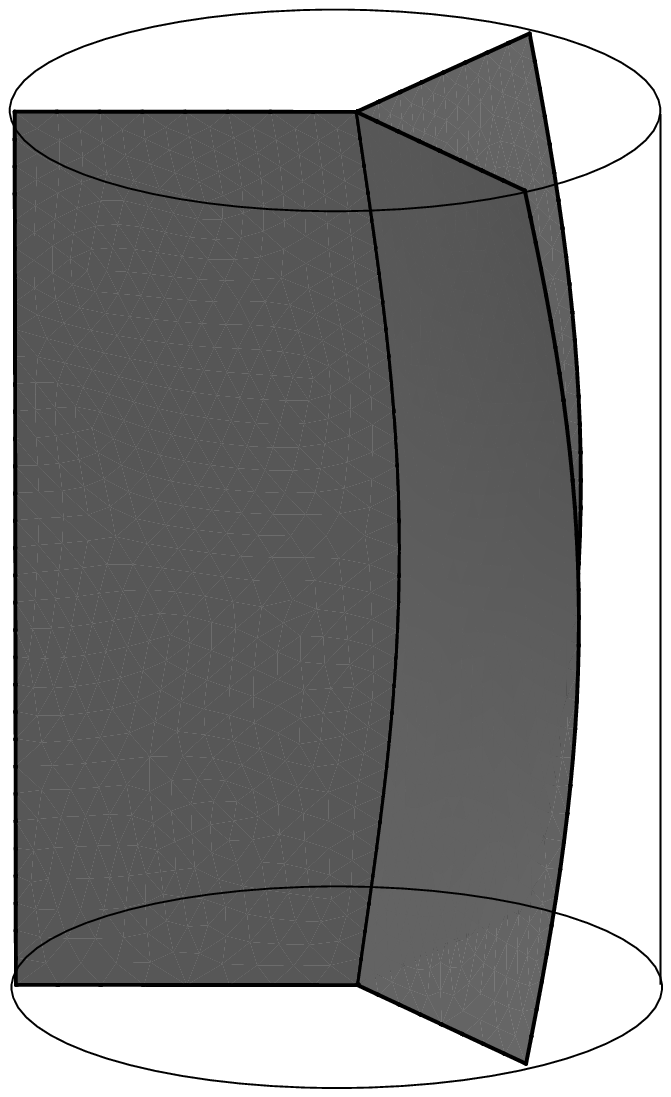}
\includegraphics[width=0.35\textwidth]{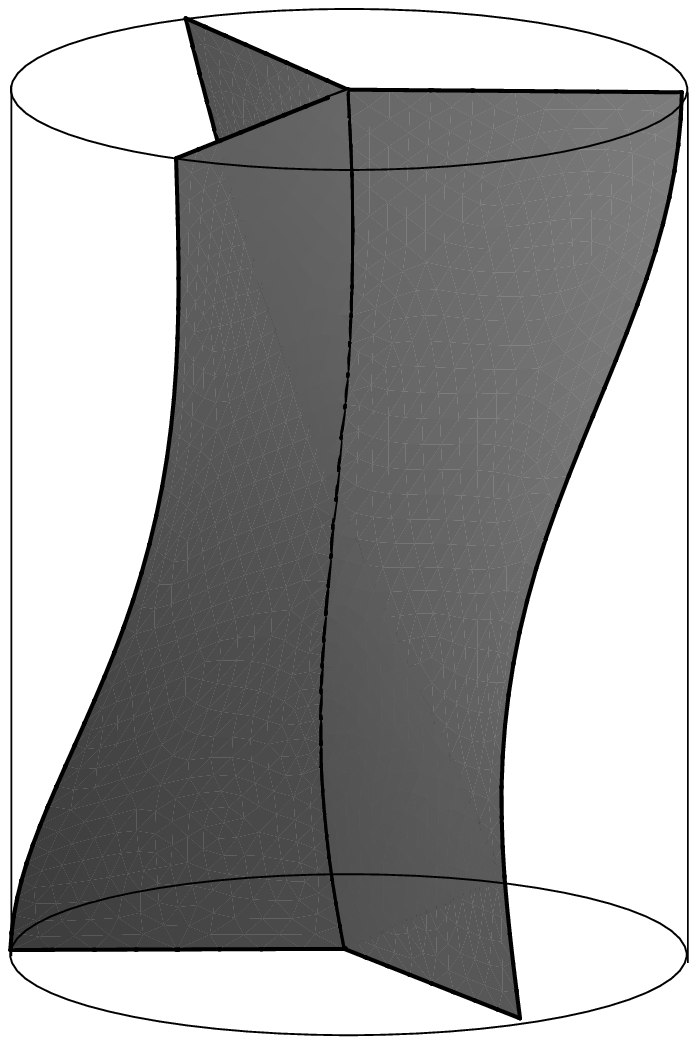}
}
\caption{Simulations of the three-film configuration: (a) Mode of instability under stretching ($\theta = 0$). (b) Mode of instability under twisting for $\theta = 60^\circ$.}
\label{fig:three}
\end{figure}

Three-fold films were examined in the same way as in the experiments: we placed a Y-shaped constraint at each end of the cylinder, with three films between them, and slowly pulled and/or rotated the constraint at one end of the film. Each film is again tessellated with about 2000 triangles. The bowing of the central Plateau border observed in the experiments during the collapse of the films is repeated in the simulations.
Figure \ref{fig:three} shows how this happens:  under stretching two of the three films distort in the same way that the single film does (Figure \ref{fig:two}(a)) while under twisting all films respond in the same way and the central Plateau border becomes distorted and no longer follows the axis of the cylinder.

The critical aspect ratios and twist angles, shown in Figure  \ref{fig:plot}, are almost indistinguishable from the simulations of the single film, indicating that the number of films present does not affect the triggering of the instability.

The discrepancy between the experimental and simulation results may be due to
the finite size of the wires and the ring supporting the upper wires, which cause deviations in the shape of the films 
particularly close to the cylinder, and the effects of gravity-driven liquid 
drainage, which causes the films to be thicker towards their base.
Both effects are more significant at the highest twist angles.

Twisting any of the simulated soap film configurations up to an angle of $180^\circ$ leads to instability at a very small height, but note that $180^\circ$ is not an upper limit to the possible twist angle. We are able to twist the three-fold film up to $192^\circ$.

Plateau's rules preclude the possibility of more than three films meeting along a line. Four films, for example, would undergo an immediate topological transition, introducing a new rectangular film between two parallel triple lines. In the case of elastic membranes, there is no such restriction, and to illustrate that this instability is not limited to soap films, we simulated an arrangement of four ``fins'' (Figure \ref{fig:one}(c)). This gives the same critical values for aspect ratio and twist angle (Figure  \ref{fig:plot}).

\section{Theory} 

We explain this effect by examining the way in which the single rectangular film collapses. It could be expected that perturbing the film would  induce an increase in area, but the presence of the cylinder walls allows the film to reduce its area due to this perturbation. Beyond a critical aspect ratio, the increase in film area due to the perturbation is more than compensated for by the reduction in area as the film slides round the walls of the cylinder at its midpoint (Figure  \ref{fig:two}(b)).

To estimate the critical aspect ratio, we will assume a certain form for the deformed shape using the coordinate system defined in Figure \ref{fig:one}(a), with $y$ a coordinate perpendicular to the planar, undeformed, film,  $x$ parallel to the anchoring wires at the ends, and $z$ vertically upwards. 
Any soap film without a pressure difference across it is a surface of zero mean curvature, so that $y(x,z)$ should satisfy Laplace's equation approximately.
We therefore assume the perturbation $y(x,z) = \varepsilon \sin(n \pi z/H) \cosh (n \pi x/H)$, where 
$\varepsilon \ll 1$ is a small parameter and the positive integer $n$ labels the possible modes of 
instability,  with wavelengths $2H / n$.  

This form for $y(x,z)$ ignores the condition that the film should meet the cylinder walls at $90^\circ$, although in their study of a structural transition of an ordered foam in a cylinder \citeauthor{reinelt01} (\citeyear{reinelt01}) noted that this condition, while the hardest to solve for, did not affect their results at small aspect ratios. 

The area of the film can then be written
\begin{equation}
A   =  2 \int_0^H {\rm d}z \int_0^{\sqrt{R^2-y^2}} {\rm d}x \sqrt{1+\left(\frac{{\rm d}y}{{\rm d}x}\right)^2+\left(\frac{{\rm d}y}{{\rm d}z}\right)^2},
\label{eq:theory1}
\end{equation}
with the upper limit of the $x$ integration representing the lateral pinching of the film due to the cylinder walls. To second order in $\varepsilon$ this becomes
\begin{eqnarray}
A  & \approx & 2 \int_0^H {\rm d}z \int_0^{\sqrt{R^2-y(R,z)^2}} {\rm d}x  +  \nonumber \\
 & & \quad	 \varepsilon^2 \left(\frac{n \pi}{H} \right)^2 \int_0^H {\rm d}z \int_0^{R} {\rm d}x  
	\left[ \sin^2 \left( \frac{n \pi z}{H} \right)  \sinh^2 \left( \frac{n \pi x}{H} \right) 
	+ \cos^2 \left( \frac{n \pi z}{H} \right)  \cosh^2 \left( \frac{n \pi x}{H} \right) \right] \nonumber \\
 & \approx &  2R \int_0^H {\rm d}z \left[ 1 - \frac{\varepsilon^2}{2 R^2} \sin^2 \left( \frac{n \pi z}{H} \right) \cosh^2 \left( \frac{n \pi R}{H} \right) \right] + \nonumber \\
 & & \quad	 \varepsilon^2 \left(\frac{n \pi}{H} \right)^2 \left[
	\int_0^H \sin^2 \left( \frac{n \pi z}{H} \right) {\rm d}z \int_0^{R}  \sinh^2 \left( \frac{n \pi x}{H} \right)  {\rm d}x + \right. \nonumber \\
 & & \quad \qquad \qquad	\left. \int_0^H \cos^2 \left( \frac{n \pi z}{H} \right) {\rm d}z \int_0^{R} \cosh^2 \left( \frac{n \pi x}{H} \right)  {\rm d}x  
	\right] \nonumber \\
 & \approx & 2 R H  - \varepsilon^2 \frac{H}{2R} \cosh^2\left( \frac{n \pi R}{H} \right) + 
	\varepsilon^2 \left(\frac{n \pi}{H} \right)^2 \left[
	\frac{H^2}{4 \pi n} \sinh\left( \frac{2 n \pi R}{H} \right) \right] \nonumber \\
 & = & 2 R H + \frac{1}{2} \varepsilon^2 \cosh\left( \frac{n \pi R}{H} \right) 
	\left[ n \pi  \sinh\left( \frac{n \pi R}{H} \right) - \frac{H}{R} \cosh\left( \frac{n \pi R}{H} \right) \right].
\end{eqnarray}

The term multiplying $\varepsilon^2$ shows the balance between the increase of surface area due to the perturbation and the decrease of surface area due to the curvature of the cylinder wall; it goes from positive to negative, indicating a reduction in area and therefore the initiation of the instability, when 
\begin{equation}
\frac{n \pi R}{H}  \tanh\left( \frac{n \pi R}{H} \right) = 1.
\end{equation}
This equation has solution $\alpha = H/R = 2.619 n $. The fundamental mode of instability, with $n = 1$, is the first to be triggered as the aspect ratio increases, giving $\alpha = 2.619$, shown as an asterisk on Figure \ref{fig:plot} and within the error bars of the simulations. 


\section{Conclusions}

We have demonstrated that any soap film or flexible membrane, such as can be found in foams and biological tissues, is susceptible to an instability that leads to its breakdown when stretched and/or twisted in a cylindrical tube. Simulations, experiments on soap films, and a simple theoretical analysis all confirm the critical value of the aspect ratio: with zero twist the critical aspect ratio of the film is $\alpha = H/R \le 2.62$, and this threshold decreases with increasing twist. This implies that there is a fundamental limit to the size of an area-minimizing film or membrane that can exist in a cylinder. Thus, for example, a tube or capillary of diameter $10\mu$m cannot support a soap film or a planar membrane along its axis of greater than about $16\mu$m in length.

That the simulations are in broad agreement with the experiments suggests that the liquid present in 
the central Plateau border and in the meniscus around the films in the experiments 
has little effect on the triggering of the instability. Simulations of a three-fold film 
arrangement with a central Plateau border of finite size suggest that the presence of 
liquid has little effect on its stability. 
Another avenue
for further research is the effect on this instability of the central 
Plateau border having a negative line tension \cite{kernw03} or bending stiffness: 
do these only affect the dynamics of collapse or do they change the point of instability itself?

Our theoretical argument for the critical aspect ratio is straightforward to 
extend to an elliptical tube in the case of pure stretching, by changing the 
upper limit in the integral in eq. (\ref{eq:theory1}), but predicting the 
critical twist angles again presents a challenging problem for future work.

Finally, we note that if the region between the single film and the cylinder wall
has a volume constraint, i.e. bubbles (or cells) are present, then this could be 
accommodated by considering the perturbation 
$y = \varepsilon \sin(2 n \pi z/H) \cosh (2 n \pi z/H)$, resulting in {\em twice} the critical aspect ratio, 
$\alpha = H/R = 5.24$, and therefore greater stability.  Note that this argument 
is related to the well-known Rayleigh-Plateau instability of fluid jets, which can also 
be couched in terms of area minimization \cite{isenberg92}.

\section*{Acknowledgements} We are grateful to the late Manuel Fortes for suggesting this problem. We thank D. Binding, F. Graner, A. Korobkin, G. Mishuris, A. Mughal, D. Weaire and S. Wilson for discussions and suggestions, EPSRC (EP/D071127/1)  and the EPSRC-P\&G strategic partnership (EP/F000049/1) for funding, and the anonymous referees for useful comments.



\end{document}